\documentclass[apj]{emulateapj}

\newcommand{\kev}{keV}
\newcommand{\etal}{et al.}
\newcommand{\nh}{$N_{\mathrm{H}}$}

\newcommand{\spitzer}{\textit{Spitzer}}
\newcommand{\microjy}{$\mu$Jy}

\newcommand{\sfr}{M$_{\odot}$~yr$^{-1}$}

\shorttitle{Obscured AGN and Starburst Disks}
\shortauthors{Ballantyne}

\usepackage{times}

\begin{document}

\title{Obscuring Active Galactic Nuclei with Nuclear Starburst Disks}


\author{D. R. Ballantyne}
\affil{Department of Physics, The University of Arizona, 1118 East 4th
  Street, Tucson, AZ 85721; drb@physics.arizona.edu}

\begin{abstract}
We assess the potential of nuclear starburst disks to obscure the Seyfert-like
AGN that dominate the hard X-ray background at $z \sim 1$. Over
1200 starburst disk models, based on the theory developed by Thompson
\etal, are calculated for five input parameters: the black hole mass,
the radial size of the starburst disk, the dust-to-gas ratio, the
efficiency of angular momentum transport in the disk, and the gas
fraction at the outer disk radius. We find that a large dust-to-gas
ratio, a relatively small starburst disk, a significant gas mass
fraction, and efficient angular momentum transport are all important
to produce a starburst disk that can potentially obscure an AGN. The
typical maximum star-formation rate in the disks is $\sim
10$~\sfr. Assuming no mass-loss due to outflows, the starburst disks
feed gas onto the black hole at rates sufficient to produce hard
X-ray luminosities of $10^{43}$--$10^{44}$~erg~s$^{-1}$. The starburst
disks themselves should be detectable at mid-infrared and radio
wavelengths; at $z=0.8$, the predicted fluxes are $\sim 1$~mJy at
24\micron\ and $\sim 10$--$30$~$\mu$Jy at 1.4~GHz. Thus, we predict a
large fraction of radio/X-ray matches in future deep radio
surveys. Unfortunately, both the 24\micron\ and radio fluxes are
comparable to those expected from the central AGN. In contrast, the
starburst disks should be easily distinguished from AGN in future
100\micron\ surveys by the \textit{Herschel Space Observatory} with
expected fluxes of $\sim 5$~mJy. Any AGN-obscuring starbursts will be
associated with hot dust, independent of AGN heating, resulting in
observable signatures for separating galactic and nuclear
star-formation. This may be an explanation for the small observed $L_{2-10\
\mathrm{keV}}/\nu L_{\nu} (6\ \mu \mathrm{m})$ ratios observed from
both $z \sim 0$ and $z \sim 1$ AGN. Finally, because of the
competition between gas and star-formation, nuclear starbursts will be
associated with lower-luminosity AGN. Thus, this phenomenon is a
natural explanation for the observed decrease in the fraction of
obscured AGN with luminosity.
\end{abstract}


\section{Introduction}
\label{sect:intro}
All Active Galactic Nuclei (AGN) are powered by accretion onto a
 central supermassive black hole
 \citep{lb69,ss73,pri81,balbus03}. However, the observational
 manifestation of this accretion can vary dramatically from object to
 object. For example, the spectral energy distributions (SED) of
 rapidly accreting quasars \citep{elv94} are very different from
 the Galactic Center source Sgr A* \citep{mf01}, or the
 black hole at the center of M87 \citep{ho99}. In this case, the
 different observational properties are a result of the two distinct
 mechanisms through which 
 energy is liberated in radiatively efficient (for the quasars) and
 inefficient (for Sgr A* and M87) accretion flows. Another important
 parameter is the orientation of the black hole-accretion disk system
 to the line of sight. As a result, blazars, AGN that are viewed down
 the axis of a radio jet, show significantly different SEDs and
 variability properties than all other AGN \citep{bach07}. Finally,
 obscuration along the line-of-sight is another key parameter in
 determining the observational characteristics of an AGN. The vast
 majority of local optically-selected AGN show evidence for being obscured
 \citep[e.g.,][]{mr95,hfs97}, and these are called Type 2 AGN, while
 the unobscured sources are defined as Type 1 AGN. Observationally,
 the obscuration is revealed as significant soft X-ray absorption
 \citep[e.g.,][]{toz06}, the disappearance of the optical broad
 permitted lines \citep{kw74}, and a reddened continuum
 \citep{wilkes05}.

Despite its near ubiquity, the origin of the obscuration around AGN is
still unclear. Infrared emission from dust near the sublimation
temperature \citep[e.g.,][]{rie78,neu79,bar87,san89} and variability
in the X-ray absorbing column density \citep{ris02,ris05} both point
to a location $\sim$1~pc from the central engine, although it is
difficult to rule out contributions from the host galaxy at larger
scales \citep{mr95,bwm03,rig06}. It is also possible that the origin
of the obscuration may arise through a variety of processes for AGN
with different luminosities or at different points in their evolution
\citep{bal06b}.

Perhaps the most significant challenge facing the pc-scale absorber (or
obscuring `torus' in the AGN unification model; \citealt{ant93}) is to account for the
observed 4-to-1 Type 2/Type 1 ratio of AGN required to fit the hard X-ray background \citep{gch07}. This fact requires the obscuring material to cover
$\sim$80\% of the sky as observed from the central black hole, and
thus have a structure so that its vertical scale height is of the
same magnitude as its radius, i.e. $H/r \sim 1$. \citet{kro07}
and \citet{sk08} have argued that the AGN emission absorbed and
re-radiated in the infrared by dust in the torus can provide enough pressure support to
inflate a structure to the required scale. Similarly, \citet{cqm07} suggested
that X-ray heating of the outer accretion disk may be sufficient to
result in a $H/r \sim 1$ geometry. Another explanation, and the one
considered here in more detail, is feedback from a nuclear starburst
disk \citep{fab98,wn02,tqm05}.

There are many reasons to consider starburst disks as a source of
obscuration in AGN. At the most basic level, gas to fuel the black hole must be
funneled toward the center of the galaxy and is therefore likely to
cause star formation en route. In fact, recent adaptive optics observations by
\citet{davies07} have shown the presence of a post-starburst population at parsec
scales in 9 local Seyfert galaxies, indicating that this may indeed be
a common phenomenon. In addition, recent models of AGN tori have
suggested that the obscuring region may be clumpy
\citep{nie02,dvb05,honig06}, a common trait of star-forming regions in starburst
galaxies \citep{fs01,skv07}. Type 2 AGN also seem to be more
correlated with star-forming activity in general than the unobscured
Type 1 objects \citep{cft95,gdhl01,lwh01,khi06,lacy07}. There is also
evidence for a correlation between the observed intensity of
star-formation in the host galaxy and the AGN nuclear luminosity \citep{shi07,wki08}.

A more profound aspect of star-formation as a source of AGN
obscuration is that it provides a connection between the black
hole-accretion disk environment and the host galaxy
\citep[e.g.,][]{kw08}. It is now accepted that the correlations
between the black hole mass and bulge properties in early-type
galaxies \citep{mag98,fm00,geb00,tre02} are evidence for a significant
connection between the growth of the black hole and the build up of a
galactic bulge
\citep[e.g.,][]{sr98,fab99,kh00,wl03,mqt05,dsh05,fvg08}. Therefore, it
is expected that as the black hole and galaxy bulge are growing, there
will be significant accretion on to the black hole that is obscured by
star formation in the host galaxy \citep[e.g.,][]{hop05}. Indeed, the
hard X-ray background, emitted by accreting black holes throughout the
history of the Universe, has a very hard spectrum, indicative of being
dominated by obscured AGN \citep{sw89,dm04,gch07}. Interestingly, the
redshift distribution of the obscured AGN that dominate the X-ray
background peaks at $z \sim 1$, very similar to the peak in the cosmic
star-formation history \citep{toz01,bar02,hopk04,bar05}. This result may imply a
connection between the obscuring material and the evolution of the
host galaxy \citep{fab98,fra99,bem06}. In fact, there is tentative
evidence that the fraction of obscured AGN does increase with $z$
\citep{bem06,tu06}. Thus, if star-forming disks do provide a
significant amount of AGN obscuration at $z \sim 1$ (where the X-ray
background sources are most common) then by studying the properties
and evolution of the absorbing material, we can directly probe the
evolution of the underlying host galaxy.

In this paper, we make use of the analytic starburst disk models
developed by \citet{tqm05} to investigate the properties of starburst
disks as a source of obscuring material in the Type~2 AGN found at $z
\sim 1$. In the paper by \citet{tqm05}, the models were able to
successfully describe many of the observed proprieties of
Ultra-Luminous Infrared Galaxies (ULIRGs), and it was noted that
under certain conditions, the photosphere of the starburst disks may reach $H/r \sim
1$. Here, we adapt the model to consider the much less intense
star-forming disks that might be expected to be found around the
Seyfert-like luminosity AGN that dominate the population at these redshifts
\citep{ueda03,bar05}. We search for the range of physical parameters
these starburst disks must have in order for them to provide
significant obscuration. Both radio and far-IR fluxes are predicted to
determine if the disks can be (or
have already been) detected by current observational
surveys. The next section provides a brief review of the \citet{tqm05}
models, and then describes how the theory was altered and applied to
this problem. Section~\ref{sect:results} presents the results of our
calculations, and describes both the physical and observational
properties of starburst disks that may obscure an average AGN. The
results are discussed in Section~\ref{sect:discuss} and we present
our conclusions in Section~\ref{sect:concl}. When necessary, the
following $\Lambda$-dominated cosmology is assumed in this paper:
$H_0=70$~km~s$^{-1}$~Mpc$^{-1}$, $\Omega_{\Lambda}=0.7$, and
$\Omega_{m}=0.3$ \citep{spe03}.

\section{Models of Nuclear Starburst Disks}
\label{sect:sbdisks}

\subsection{Review of Starburst Disk Models}
\label{sub:review}
The properties of the nuclear starburst disks are calculated using the
  one-dimensional analytic theory developed by \citet{tqm05}, and we
  solve the equations presented in their Appendix~C. Here, we will
  briefly review the main assumptions and ideas behind the theory, and
  refer readers to the paper by \citet{tqm05} for a complete
  discussion.

The starburst disks are calculated as a single phase medium at a
radius $r$ from the center of a galaxy with a velocity dispersion
$\sigma$ and a supermassive black hole with mass
$M_{\mathrm{BH}}$. Further, the disk is assumed to be rotating at the
Keplerian rate at all radii, and the gravitational potential is the
sum of the point-mass of the black hole and an isothermal
sphere describing the galactic bulge. The star-formation in the disk
is a local phenomenon and is assumed to always adjust so that Toomre's
$Q$ parameter is unity; that is, $Q=\kappa_{\Omega} c_{s}/\pi G
\Sigma_{\mathrm{g}}=1,$ where $\kappa_{\Omega}$ is the epicycle
frequency, $\Omega$ is the Keplerian frequency, $c_{s}$ is the sound speed and
$\Sigma_{\mathrm{g}}$ is the gas surface density. This condition on
$Q$ allows the gas density $\rho$ at any radius to be immediately written as
a function of $\Omega(r)$. The vertical support
of the disk is the infrared (IR) radiation pressure of the starburst
against dust (the UV optical depth is always much greater than unity
in these disks, so the vast majority of the radiation is re-radiated
in the mid- and far-IR). Therefore, these starbursts are locally
radiating at their Eddington limit. The vertical optical depth of the disk in the
IR is $\tau = \Sigma_{\mathrm{g}} \kappa / 2$, where $\kappa(T,\rho)$ is the
Rosseland mean opacity.

Fuel is fed to the outer radius of the disk at a rate of
$\dot{M}_{\mathrm{out}}$. An unspecified global torque acts on the
disk allowing the removal of angular momentum and gas to accrete slowly
inwards at a rate $\dot{M}$. This torque is assumed to allow the
radial velocity of the gas to be equal to a constant fraction $m$ of
the local sound speed; thus, $\dot{M}=2\pi r \Sigma_{g} m
c_s$. However, gas can be removed from the disk by local star-formation,
so that the accretion rate through the disk can also be written as $\dot{M} =
\dot{M}_{\mathrm{out}} - \int^{R_{\mathrm{out}}}_{r} 2\pi r^{\prime}
\dot{\Sigma}_{\ast} dr^{\prime}$, where $\dot{\Sigma}_{\ast}$ is the
star-formation rate (SFR) per unit area. Therefore, there is a competition
between star-formation using up the gas in the disk and the torque
attempting to transport material through the disk. As discussed in
\citet{tqm05}, this can be understood by considering the advection
timescale through the disk, $\tau_{\mathrm{adv}}=r/v_{r}$, and the
star-formation timescale, $\tau_{\ast}=1/(\eta \Omega)$, where $v_{r}$
is the radial velocity of the gas, and $\eta$ is the star-formation
efficiency. If $\tau_{\mathrm{adv}} < \tau_{\ast}$ then material can
move through the disk quickly enough that it can fuel a black hole while
forming stars. \citet{tqm05} showed that by equating these
timescales at the outer radius $R_{\mathrm{out}}$, one derives a critical accretion rate,
$\dot{M}_{\mathrm{crit}}$, below which nearly all the gas is converted into stars at
large radii.

The model disks are calculated from $R_{\mathrm{out}}$ inwards until a
point is reached where dissipation by disk accretion dominates the
heating over star-formation. At this point, $Q = 1$ can no longer be
maintained and the starburst disk structure calculation is
terminated. The actual value of the inner radius depends on the
particular parameters of the model, but is typically $< 1$~pc.

\subsection{A Survey of Starbursts}
\label{sub:seyferts}
\citet{tqm05} calculated starburst disk structures appropriate for
ULIRGs, but the X-ray background is dominated by AGN whose
host galaxies must have, on average, SFRs much lower than ULIRGs
\citep{bp07}. Thus, we have to calculate new starburst models for this
different scenario. Since it is clear that it is possible for
starburst disks to potentially obscure an AGN \citep{tqm05}, these new
models are designed to cover a wide range of parameter space to
determine the variety of properties that AGN-obscuring starburst disks
may have.

Each starburst disk calculation has 5 input parameters that were
varied through different values: $\log(M_{\mathrm{BH}}/M_{\odot})=7$,
$7.5$, $8$, $8.5$ (the galactic velocity dispersion is then calculated
through the $M_{\mathrm{BH}}$-$\sigma$ relation:
$M_{\mathrm{BH}}=2\times 10^8 \sigma^{4}$; e.g., \citealt{tre02});
$m=0.0075$, $0.01$, $0.025$, $0.05$, $0.075$, $0.1$, $0.2$;
$R_{\mathrm{out}}=50$, $100$, $150$, $200$, $250$~pc;
$f_{\mathrm{gas}}=0.1$, $0.5$, $0.9$ (the gas fraction determines
$\dot{M}_{\mathrm{out}}$ and decreases to smaller $r$;
\citealt{tqm05}); and the dust-to-gas ratio of the gas (either
$1\times$, $5\times$ or $10\times$ the local ISM value). As emphasized
by \citet{tqm05}, the starburst disk structure is very sensitive to
the temperature dependence of the Rosseland mean opacity. Similarly,
the overall opacity value is an important parameter, as increasing the
opacity can result in a larger temperature without necessarily
enhancing the SFR. There are numerous observational hints that the
region around an AGN may have enhanced metallicity
\citep{hf99,nag06,arav07}, as well as increased grain growth
(\citealt{cqm07} and references therein). Therefore, it is important
to calculate starburst disks with differing dust-to-gas
ratios. Following \citet{tqm05}, we use the Rosseland mean opacity
curve calculated by \citet{sem03} with the following parameters: a
density of $10^6$~cm$^{-3}$, ``normal'' silicates, homogeneous
distribution of dust materials, and spherical particles. To determine
the opacity at different dust-to-gas ratios, this basic curve,
appropriate for typical ISM values, was multiplied by the desired
ratio \citep{ferg07}.

\begin{figure}
\epsscale{1.2}
\plotone{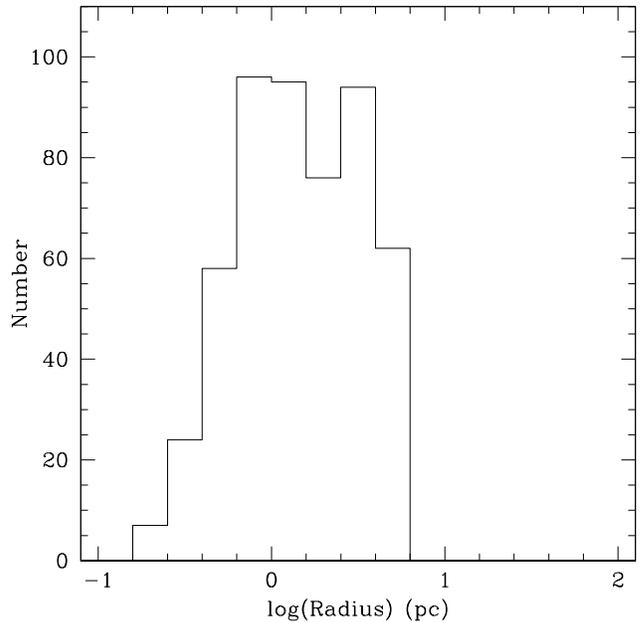}
\caption{Histogram of the radius where the SFR reaches
  its maximum. The models making up the histogram were selected so
  that the maximum star-formation rate occurred in the high temperature
  dust sublimation region that satisfied the two criteria described
  in the text. Of the 1260 starburst disk models calculated, a
  total of 512, or $41$\%, satisfy these conditions. The peak
  star-formation rate always occurs at $r \ga 0.2$~pc, with $r \approx
  1$ and $3$~pc being the two most common values.}
\label{fig:radiushisto}
\end{figure}

Starburst disk models were run for each permutation of the input
parameters resulting in a total of 1260 distinct starburst disk
structures. However, before analyzing the properties of these disk
models, we must select the models that are most likely to both fuel and obscure a
central active nucleus (i.e., ones where $\tau_{\mathrm{adv}} <
\tau_{\ast}$). The potential for obscuring an AGN arises when the
central temperature of the starburst disk surpasses the sublimation
temperature of the dust \citep{tqm05}. At that point, the opacity of the
gas drops precipitously and the SFR rises dramatically
in order to keep $Q=1$. In this situation, the effective temperature
of the disk (roughly the surface temperature) may remain below the
sublimation temperature, leading to a significant vertical opacity
gradient in the disk. \citet{tqm05} argued that, although the disk may
remain thin with a $H/r < 1$, the photosphere of the disk could puff up to $H/r \sim
1$. Therefore, each of the 1260 models was searched to determine if
there existed a region that satisfied both of the following conditions:
\begin{enumerate}
\item the central temperature was greater than 900~K,
\item the SFR was greater than 10\% of the SFR at $R_{\mathrm{out}}$.
\end{enumerate}
The second criterion was necessary to exclude a small number of models
that marginally exceeded the temperature condition, but did not reach
the sublimation temperature. Only 512 models, or $41$\% of the
calculations, produced an inner starburst that satisfied both of the
imposed conditions. A histogram of the radius at which the SFR reached
its maximum within this region is shown in Figure~\ref{fig:radiushisto}.
 The histogram shows
 that the most common radius to find this burst of star-formation is $r
 \approx 1$--$3$~pc. Such radii are very similar to the scales of the
 obscuring medium measured in nearby AGN \citep{jaffe04,tris07}. Thus,
 these starburst disk models reproduce the size scales necessary to
 obscure AGN. It is worth emphasizing that in choosing the range of
 model parameters, we made no \textit{a priori} assumptions about what values
 would be more (or less) likely to produce these pc-scale
 starbursts, but only considered values to be typical of observed
 Seyfert-like galaxies. The distributions of observables shown below are derived
 solely from the models shown in Fig.~\ref{fig:radiushisto}, which we
 consider to be those that are  most likely to obscure the central AGN.

\section{Results}
\label{sect:results}
\subsection{Physical Properties of the Starburst Disks}
\label{sub:physical}
\begin{figure}
\epsscale{1.2}
\plotone{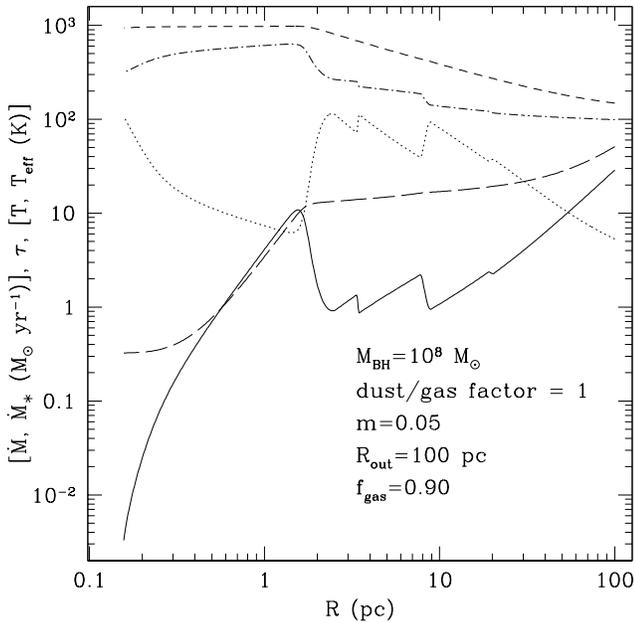}
\caption{Plot of the mass accretion rate $\dot{M}$ (long-dashed line;
  in M$_{\odot}$~yr$^{-1}$), the SFR
  $\dot{M}_{\ast}$ (solid line; in M$_{\odot}$~yr$^{-1}$), the
  vertical optical depth $\tau$ (dotted line), the central
  temperature $T$ (short-dashed line; in K), and the effective temperature
  $T_{\mathrm{eff}}$ (dot-dashed line; in K), all as a function of
  radius. The data are drawn from a starburst disk model with the
  following parameters: $M_{\mathrm{BH}}=10^8$~M$_{\odot}$,
  dust-to-gas factor$=1$, $m=0.05$, $R_{\mathrm{out}}=100$~pc and
  $f_{\mathrm{gas}}=0.9$.}
\label{fig:example}
\end{figure}

As a guide to the physical structure of one of the starburst disks,
Figure~\ref{fig:example} plots the radial dependence of several
interesting variables from one representative model.
The details behind the models are described by \citet{tqm05}, but it
is worth emphasizing here that star-formation occurs throughout the
disk (solid line), from $R_{\mathrm{out}}=100$~pc to the inner-radius at $\la
1$~pc. However, the central temperature of the disk (short-dashed
line) increases to smaller radii as the rising density causes the disk
to become more optically-thick to the stellar radiation. Eventually,
this temperature reaches the sublimation temperature of the dust, and
the opacity drops (dotted line), resulting in a significant burst of
star-formation. Since the effective temperature (dot-dashed line) is
still below this sublimation temperature, a significant vertical
opacity gradient will exist in this region. Thus, this is the key
region for where star-formation can puff up the disk photosphere to $H/r \sim 1$,
and not at larger radii, even if the SFRs are comparable. This inner
burst of star-formation nearly exhausts the gas supply in these
models, but a small amount survives to make its way onto the inner
accretion disk to feed the black hole (long-dashed line). 

\begin{figure*}
\begin{center}
\includegraphics[angle=-90,width=0.90\textwidth]{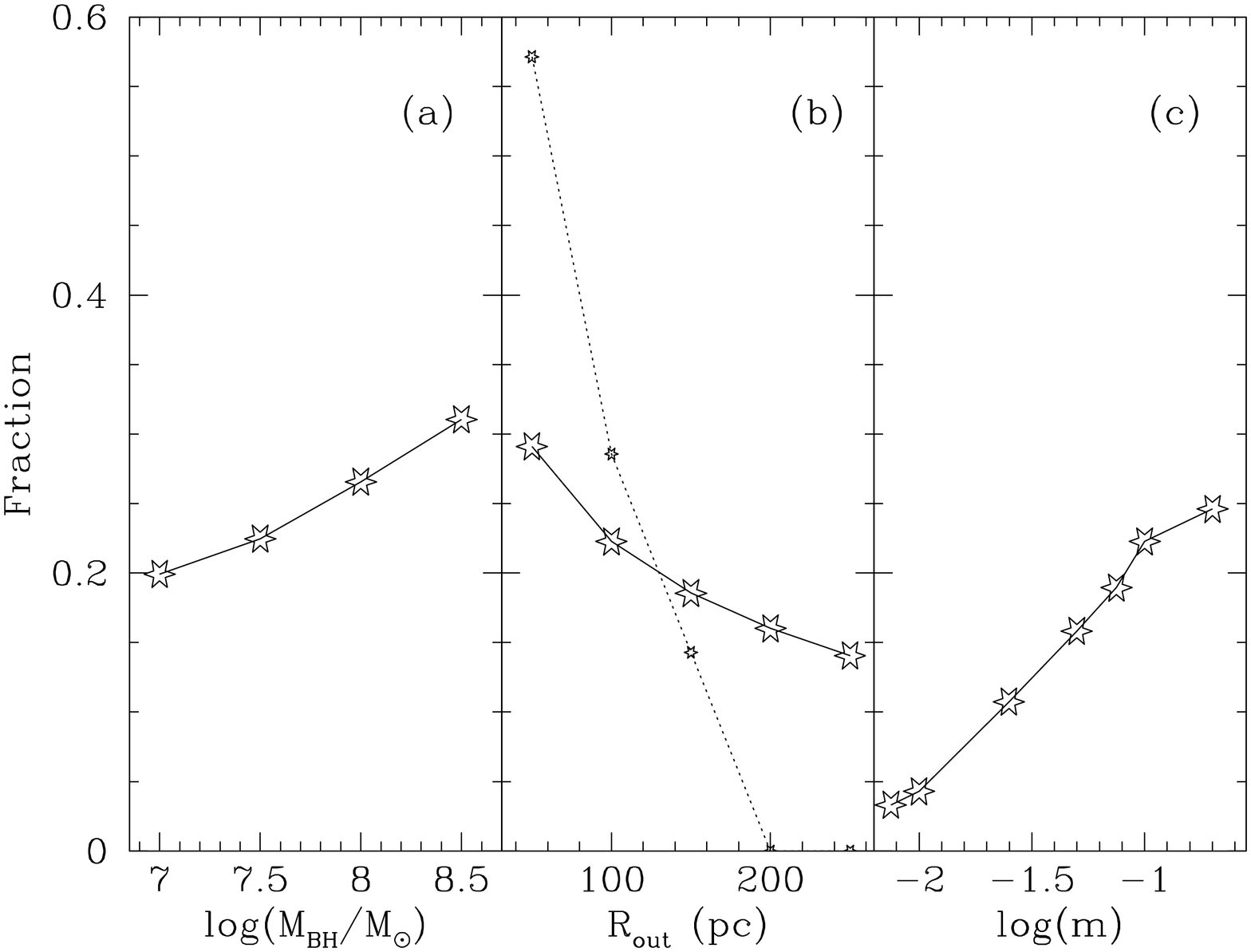}
\\
\includegraphics[angle=-90,width=0.75\textwidth]{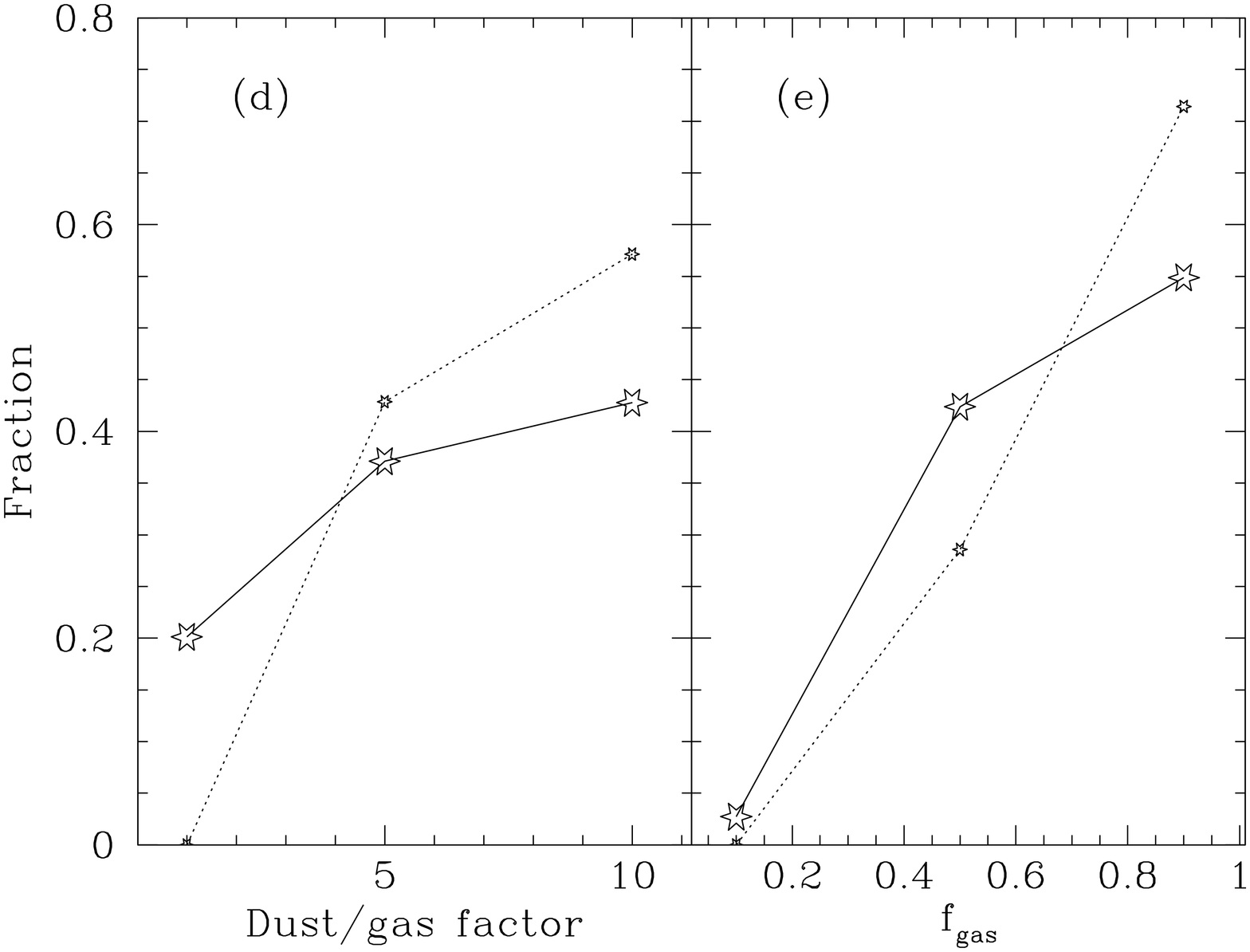}
\end{center}
\caption{The different panels show the fraction of the models
  with pc-scale bursts of star-formation (i.e., those shown in
  Fig.~\ref{fig:radiushisto}) that have a specific input
  parameter. The panels show (a) the mass of the central black hole,
  (b) the outer radius of the starburst disk, (c) the logarithm of the
  Mach number $m$, (d) the dust-to-gas
  multiplicative factor and (e) the gas fraction at
  $R_{\mathrm{out}}$. The solid lines are the results when including
  all the successful models. The dotted lines plot the fractions when
  the additional constraints of $\log(M_{\mathrm{BH}}/M_{\odot})=7$
  and $m=0.025$ are imposed.}
\label{fig:props}
\end{figure*}

We now turn to the physical properties for the ensemble of
`successful' models shown in Fig.~\ref{fig:radiushisto}. What is necessary for these disks to be able to
obscure an AGN on pc scales? The solid lines in Figure~\ref{fig:props} show the
distributions of the input parameters derived from those models that
have a pc-scale burst of star-formation. 
These plots show that successful models can be produced at every value
of the input parameters, but pc-scale starbursts are more commonly 
generated if the disks have specific properties. For example,
panel (a) shows that more massive black holes can more easily produce
such starbursts. This is perhaps unsurprising since in these $Q=1$
structures a larger black hole mass results in a denser and hence hotter starburst
disk \citep{tqm05}. In contrast, panel (b) shows that a smaller
starburst disk is more likely to have the inner ring of
star-formation. Smaller starburst disks retain a larger fraction of
their gas at a given radius than larger disks. Thus, the opacity is
greater and the temperature can more easily reach the sublimation
temperature of dust. The smallest $R_{\mathrm{out}}$ considered, 50~pc, is
an interesting number as its roughly the same size as the star-forming
regions found around nearby Seyferts by \citet{davies07}. The models considered
here do show that such small starburst disks are more likely to
produce an AGN obscuring region than larger disks, in agreement with
the observations. 

There is a similarly strong trend shown in panel (c) involving the
Mach number $m$. A larger value of $m$ reduces $\tau_{\mathrm{adv}}$,
and thus causes more efficient mass accretion through the starburst
disk. Figure~\ref{fig:props} shows that an inner burst of star
formation at pc-scales is more common at higher Mach numbers. In fact,
$\sim 50$\% of such models had
$m \geq 0.1$. As with the smaller values of $R_{\mathrm{out}}$,
efficient mass accretion results in a larger fraction of gas at
pc-scales, which increases the opacity and hence temperature of the
gas. The same reasoning can also be applied to the dust-to-gas factor
shown in panel (d), where clearly a larger dust-to-gas ratio results
in a more common inner pc-scale burst of star-formation. A enhanced
dust-to-gas ratio increases the vertical opacity to the starburst
radiation, leading to larger temperatures and thus increasing the
likelihood of surpassing the sublimation temperature. 

The gas fraction at $R_{\mathrm{out}}$ (Fig.~\ref{fig:props}(e))
is another important parameter for the existence of a pc-scale
starburst, with $f_{\mathrm{gas}}=0.1$ producing only $\sim 2$\% of
the `successful' models. The mass accretion rate at $R_{\mathrm{out}}$ is
directly related to $f_{\mathrm{gas}}$, so that a larger accretion
rate onto the outer disk is the result of increasing
$f_{\mathrm{gas}}$. Hence, it is easier to exceed the
$\dot{M}_{\mathrm{crit}}$ crucial for feeding and obscuring the black
hole. A larger
$f_{\mathrm{gas}}$ at $R_{\mathrm{out}}$ will also likely result in a
larger fraction of gas on pc-scales, and the extra heating argument
described above will hold. The accretion rate varies as
$f_{\mathrm{gas}}^2$ \citep{tqm05}, thus models with $f_{\mathrm{gas}}
\sim 0.5$ are adequate to produce potentially AGN-obscuring starburst disks. 

It is clear from the physics of the starburst model that in order to
produce this pc-scale burst of star-formation that may obscure an AGN,
the temperature of the disk must exceed the sublimation temperature of
dust producing a sudden drop of the vertical
opacity. Figure~\ref{fig:props} shows that this situation is more
likely to occur if one of the following is true: a high-mass black
hole, a smaller starburst disk, efficient angular momentum transport
in the disk, a reasonable gas mass fraction, or a large dust-to-gas
ratio. However, not all of these conditions have to be simultaneously
satisfied. For example, the dotted lines in Fig.~\ref{fig:props} show
the distribution of $R_{\mathrm{out}}$, $f_{\mathrm{gas}}$ and the
dust-to-gas factor for the subset of models where
$\log(M_{\mathrm{BH}}/M_{\odot})=7$ and $m=0.025$, two non-ideal
conditions for the existence of the inner pc-scale starburst. In this
case, the small starburst disks and large dust-to-gas ratios are
basically required in order to generate a potentially AGN-obscuring
starburst. Therefore, while it may be harder to produce a pc-scale
burst of star-formation around a smaller mass black hole, it is not
impossible if the other properties of the starburst disk are
favorable.

\subsection{Star-formation Rates}
\label{sub:sfrs}
\begin{figure}
\epsscale{1.2}
\plotone{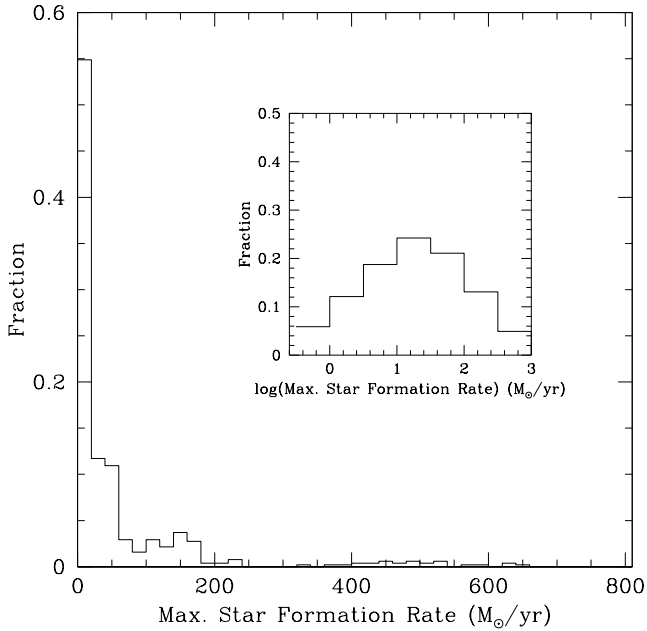}
\caption{Histogram of the maximum SFRs found in the models with
  pc-scale starbursts. About 55\% of the models have a maximum SFR$<
  20$~\sfr. The distribution shows a weak tail to high SFRs, with a
  small number of models with peak SFRs $> 300$~\sfr. The inset shows
  the same distribution, but now on a logarithmic axis. The most
  common SFRs are $10$--$30$~\sfr\, and only $\sim 5$\% of the
  `successful' calculations have SFRs $> 300$~\sfr.}
\label{fig:sfr}
\end{figure}

We now understand the conditions necessary for the starburst disk to
produce a potentially AGN obscuring ring of star-formation. Focusing
in on the star-formation events themselves, Figure~\ref{fig:sfr} plots
the distribution of maximum SFRs found in these pc-scale starbursts.
The plot shows that just over half the models have maximum SFRs $<
20$~\sfr\ (with values $\sim 10$~\sfr\ most common), rates that are
very typical of luminous IR galaxies (LIRGs; \citealt{wang06}), but much less than
those observed in ULIRGs ($\gtrsim 100$~\sfr; \citealt{rz08}). However, the
star-formation per unit area in the model bursts are very high ($\sim
10^6$~\sfr~kpc$^{-2}$), exceeding
those observed in ULIRGs ($\sim 10^3$~\sfr~kpc$^{-2}$; \citealt{gc08}). Thus, these pc-scale starbursts are
basically miniature ULIRGs: small regions of very efficient
star-formation, but because the area is so small, the maximum SFRs can
be LIRG-like. Interestingly, \citet{bp07} showed that the host galaxies of the
AGN that dominate the X-ray background must, on average, have SFRs$< 100$~\sfr\ to
avoid violating the cosmic IR background. This constraint is
consistent with the predictions of Fig.~\ref{fig:sfr}, where $\sim
80$\% of the models have maximum SFRs less than $100$~\sfr.

The distribution of SFRs shows a small number of models with
maximum SFRs $> 300$~\sfr. The Fig.~\ref{fig:sfr} inset indicates that
only 5\% of the models present such ULIRG-like SFRs. These massively
star-forming disks only occur when $\log(M/M_{\odot})=8.5$, $m=0.1$ or
$0.2$, and $f_{\mathrm{gas}} > 0.90$. Clearly, significant gas
resources are necessary to power such bursts of star-formation.

\subsection{Observational Consequences}
\label{sub:observe}
\subsubsection{AGN Properties}
\label{subsub:agn}
As mentioned in Sect.~\ref{sub:physical}, if $\tau_{\mathrm{adv}} <
\tau_{\ast}$ a small amount of gas
survives its passage through the starburst disk and will feed onto the
sub-pc accretion disk, and may ultimately accrete onto the black
hole. Under the assumption that the accretion rate at the inner-most
edge of the starburst disk is the same as the accretion onto the
black hole, the bolometric luminosity of the resulting AGN can be
estimated from $L_{\mathrm{bol}}=\eta \dot{M}c^2$, where $\eta=0.1$ is the
radiative efficiency of the accretion disk. This is not a very realistic assumption, since outflows from
the central regions of AGN are commonly observed, as well as being a
standard prediction from numerical simulations of accretion flows
\citep[e.g.,][]{ckg03,hk06}. However, these outflows ensure that the accretion rate onto
the black hole is always less than the accretion rate at the outer
edge of the accretion disk, thus the luminosities calculated from the
predicted $\dot{M}$ are strict upper-limits\footnote{The mass-loss rate from
an accretion disk depends on the location from where it is launched
and how it is driven, both extremely uncertain numbers for almost all AGN
\citep{ckg03}. Recently, \citet{mill08} argued that an outflow from the
Galactic black hole GRO J1655-40 was launched from the disk by
magnetic processes. The mass-loss rate in this case was estimated to
be $>5$--$10$\% of the accretion rate.}. A further, implicit,
assumption in this calculation is that there are no other sources of
fuel for the black hole aside from that provided by the starburst
disk.

\begin{figure}
\epsscale{1.2}
\plotone{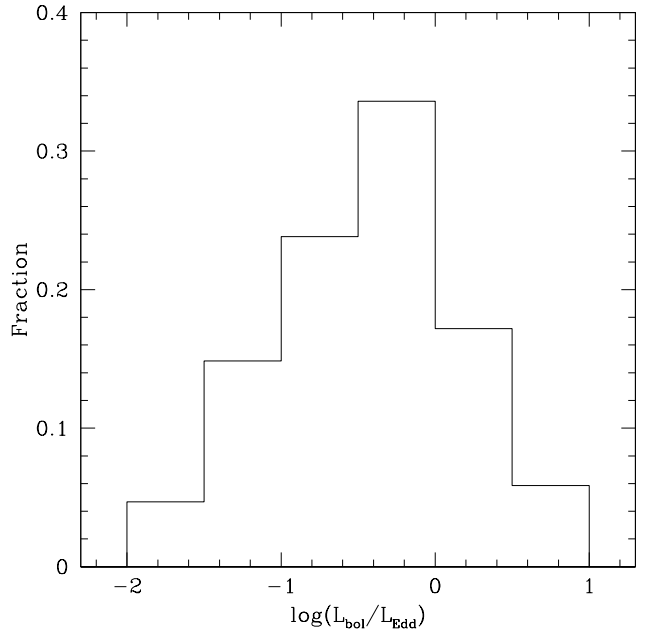}
\caption{Histogram of the Eddington ratio, or
  $L_{\mathrm{bol}}/L_{\mathrm{Edd}}$, predicted from a central AGN
  being fed from the surrounding starburst. Mass-loss in the inner
  accretion disk is not included, so these ratios are strict
  upper-limits. Sixty percent of the models have $0.1 \leq
  L_{\mathrm{bol}}/L_{\mathrm{Edd}} \leq 1$, with a further $\sim
  20$\% apparently accreting at super-Eddington rates.}
\label{fig:lratio}
\end{figure}

Keeping those assumptions in mind, we plot in Figure~\ref{fig:lratio}
the Eddington ratio $L_{\mathrm{bol}}/L_{\mathrm{Edd}}$ distribution
from the models with pc-scale starbursts. Here, $L_{\mathrm{Edd}}=4\pi
GM_{\mathrm{BH}}m_p c/\sigma_{\mathrm{T}}$ is the Eddington ratio for
  spherical accretion, where $m_p$ is the proton mass and
  $\sigma_{\mathrm{T}}$ is the Thomson cross-section.
The histogram shows a log-normal-like distribution with a peak at
$L_{\mathrm{bol}}/L_{\mathrm{Edd}} \approx 0.5$. In all, about 20\% of
the models have Eddington ratios less than 0.1, $\sim 60$\% have ratios $0.1 \leq
  L_{\mathrm{bol}}/L_{\mathrm{Edd}} \leq 1$, while $\sim 20$\% predict
  super-Eddington accretion. Since these ratios will be upper-limits, we
  find that the majority of the Eddington ratios predicted by fueling from the model
  starburst disks are consistent with the values inferred from
  obscured, $z \sim 1$ AGN \citep[e.g.][]{ballo07}.

\begin{figure}
\epsscale{1.2}
\plotone{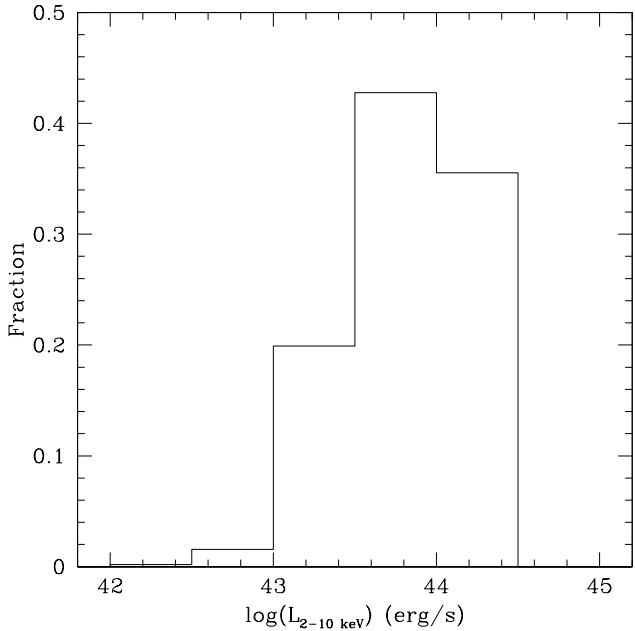}
\caption{Histogram of the 2--10~\kev\ luminosity of AGN fed from the
  starbursts disks with pc-scale bursts of star-formation. These are
  strict upper-limits. The range of luminosities is consistent with
  those AGN observed to dominate the X-ray background at $z \sim 1$
  \citep{ueda03,bar05}.}
\label{fig:lx}
\end{figure}

While the Eddington ratio is an interesting parameter, it is not a
true observable. Assuming a standard radiatively-efficient AGN
spectrum, bolometric corrections can be used to
determine the AGN luminosity in the 2--10~\kev\ X-ray band. Recently, \citet{vf07} determined the 2--10~\kev\ bolometric
correction for a number of nearby AGN with both X-ray and UV
data. These authors found that the bolometric correction may depend on
the Eddington ratio of the AGN, with a value of $L_{\mathrm{bol}}/L_X
\sim 20$ for $L_{\mathrm{bol}}/L_{\mathrm{Edd}} < 0.1$, while $L_{\mathrm{bol}}/L_X
\sim 55$ for larger values, where $L_X$ is the AGN luminosity in the
2--10~\kev\ band. Employing these bolometric corrections, strict
upper-limits to $L_X$ can be calculated for our models and its
distribution is plotted in Figure~\ref{fig:lx}.
Over $90$\% of the predicted X-ray luminosities fall between
$L_X = 10^{43}$~erg~s$^{-1}$ and 10$^{44.5}$~erg~s$^{-1}$, which is
precisely the range of luminosities observed from the AGN that
dominate the hard X-ray background at $z \sim 1$ \citep{ueda03,bar05}. It is interesting that
although the nuclear accretion process is not modeled, the AGN
accretion rates predicted by the competition between star-formation
and a global feeding process on scales of tens of pc gives such good
agreement with the observations. Although the details are unclear,
this result may support models of AGN feeding through star-forming
disks and global, bar-like instabilities (see
Sect.~\ref{sub:torques}).

\subsubsection{The Starburst Disk}
\label{subsub:sbdisk}
\begin{figure}
\epsscale{1.2}
\plotone{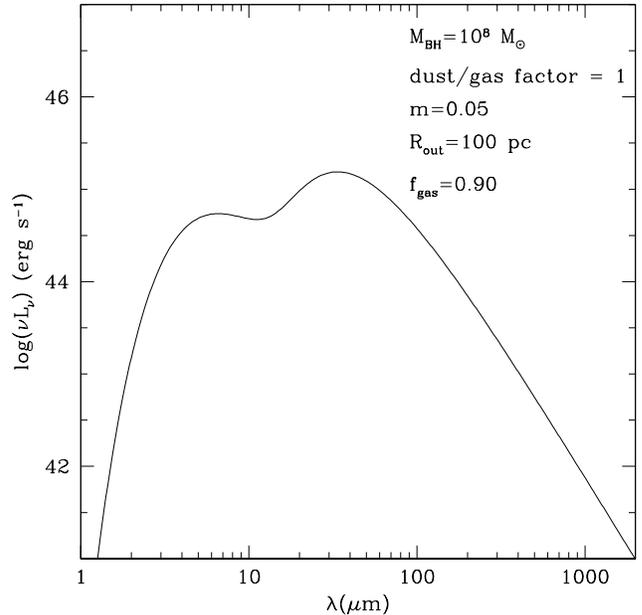}
\caption{Example SED of a starburst disk from a model with the
  following parameters:  $M_{\mathrm{BH}}=10^8$~M$_{\odot}$,
  dust-to-gas factor$=1$, $m=0.05$, $R_{\mathrm{out}}=100$~pc and
  $f_{\mathrm{gas}}=0.9$. The inner pc-scale starburst causes a bump
  in emission at $\lambda \sim 5$\micron. The SED is calculated from
  eq.~\ref{eq:sed} and assumes a face-on geometry; hence the
  luminosity is an upper-limit.}
\label{fig:sedexample}
\end{figure}

To determine the observational properties of the starburst disk
itself, we calculate a simple SED for each model assuming
blackbody emission at each radius and a face-on viewing angle \citep{tqm05}:
\begin{equation}
\lambda L_{\lambda}={2 \pi h c^2 \over \lambda^4}
\int^{R_{out}}_{R_{in}} {2 \pi r dr \over \exp[hc/\lambda
    k_{\mathrm{B}} T_{\mathrm{eff}}(r)]-1},
\label{eq:sed}
\end{equation}
where $R_{\mathrm{in}}$ is the inner radius for each model, $h$ is
Planck's constant and $k_{\mathrm{B}}$ is Boltzmann's constant. An
example SED, calculated from the same starburst model as described in
Fig.~\ref{fig:example}, is shown in Figure~\ref{fig:sedexample}.
The SED shows two maxima: one at $\sim 30$\micron\ from the
star-formation at the outer edge of the disk, and one at $\sim
5$\micron\ from the pc-scale starburst. This shorter-wavelength peak
is a generic feature of the models with the inner ring of
star-formation, and is due to the loss of dust opacity in this region
that limits the reprocessing available for the hot starburst
emission. Thus, the effective temperature at these radii moves into
the near-IR regime. As these SEDs are calculated assuming the disk is
viewed face-on, the luminosities and fluxes discussed below are
upper-limits.

\begin{figure}
\epsscale{1.2}
\plotone{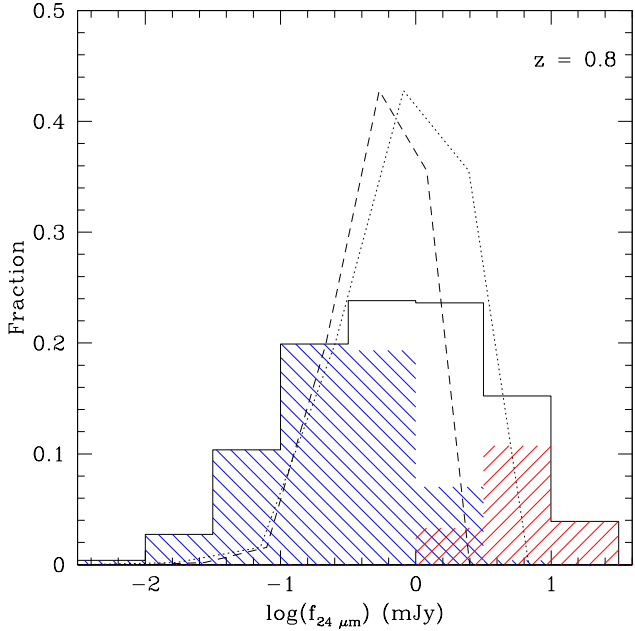}
\caption{The solid line plots the distribution of predicted 24\micron\
  flux at $z=0.8$ from models with pc-scale starbursts. The blue
  region denotes the location of models with a maximum SFR
  $\leq 30$~\sfr, while the red region shows the distribution of
  models with a maximum SFR $> 100$~\sfr. All of these fluxes are
  derived from SEDs calculated by equation~\ref{eq:sed} and are thus
  upper-limits. The dashed and dotted lines are predictions of the
  24~\micron\ flux due to AGN heating of a constant density torus at a
  distance of 1 and 10~pc, respectively \citep{bal06b}. These calculations made
  use of the distribution of X-ray luminosities from
  Fig.~\ref{fig:lx}.}
\label{fig:24um}
\end{figure}

Figure~\ref{fig:24um} plots the distribution of 24\micron\ flux at $z=0.8$
predicted from the starburst disk SEDs.
The choice of $z=0.8$ is illustrative, but is a common redshift for
the population of obscured AGN that we wish to describe
\citep[e.g.,][]{bar05}. The most
common 24~\micron\ flux predicted by the starburst models is $\sim
1$~mJy, but values can be as low as 10~$\mu$Jy, or as high as 30~mJy. As
mentioned above, these values should be treated as upper-limits. The blue
shaded region shows where the models with a maximum SFR$ \leq
30$~\sfr, while the red area plots models with a maximum SFR$> 100$~\sfr. The starburst disks with lower SFRs, make up over 50\% of
the models (Fig.~\ref{fig:sfr}) and dominate the 24~\micron\
predictions at fluxes $< 1$~mJy. For fluxes between 1--3~mJy, the models with
intermediate SFRs dominate before ceding to the small number of high
SFR disks. 

The fluxes shown in Fig.~\ref{fig:24um} are well within the sensitivity limits of many
of the deep \spitzer\ surveys over the last several years
\citep{lacy04,rig04,fra05,stern05,ah06,barm06,brand06,pg08}. Thus, if these disks exist around AGN, they should have been
detected by these surveys. However, the AGN itself can heat
dust to near- and mid-IR emitting temperatures. The calculations
presented in this paper do not include the feedback of the central AGN
on the starburst disk. To estimate the 24~\micron\ flux from AGN
heating, we make use of the calculations by \citet{bal06b} who modeled AGN SEDs
from the X-ray to the far-IR assuming a constant density torus with no
star-formation. These models allow one to easily connect the X-ray
luminosity to predicted IR fluxes, so the distribution of X-ray
luminosities shown in Fig.~\ref{fig:lx} can then be converted to a
distribution of 24\micron\ flux with the results plotted in
Fig.~\ref{fig:24um}. The dashed and dotted lines assume a distance
from the AGN to the inner radius of the torus of 1 and 10~pc,
respectively. The lines were derived from models that are averaged
over all X-ray column densities, and assume a $(1+z)^{0.3}$ evolution
of the AGN Type 2/Type 1 ratio \citep{bem06,tu06}, although this assumption does
not affect the results shown here. We see that a dust torus heated by
a distribution of AGN
X-ray luminosities from Fig.~\ref{fig:lx} results in very similar
24\micron\ fluxes as that predicted from the starburst disks. Thus, we
recover the long-standing problem of attempting to determine the
relative influence of star-formation and AGN heating in an observed
spectrum. However, the nuclear starburst models described here present
a new wrinkle: observable hot-dust emission in the near- and mid-IR
that is \textit{not} caused by AGN heating. This may mislead some
methods of AGN identification in the IR that are based on the shape of
the near- and mid-IR spectrum \citep[e.g.,][]{don07}. We conclude that the nuclear
starburst disks will be easily visible in the mid-IR, but since the
predicted fluxes are nearly identical to those expected from AGN
heating, it will be extremely difficult to positively identify them or
measure their properties (see Sect.~\ref{sect:discuss}).

\begin{figure}
\epsscale{1.2}
\plotone{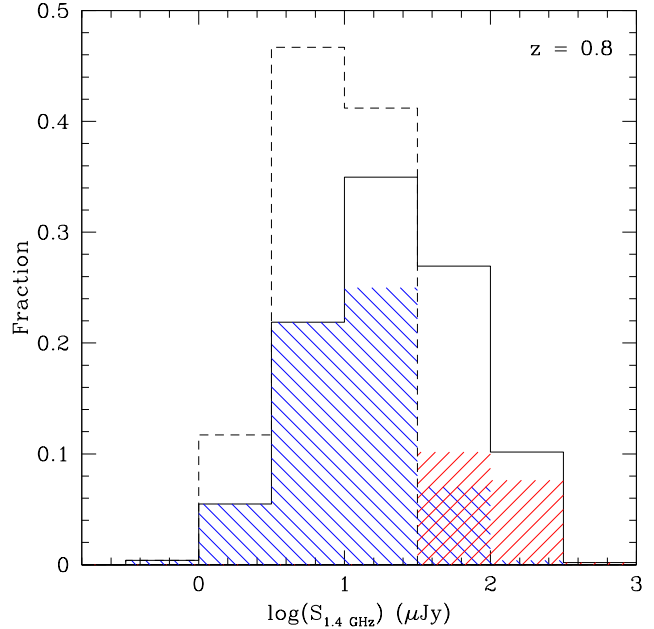}
\caption{The solid histogram plots the distribution of 1.4~GHz radio
  fluxes at $z=0.8$ as estimated from the radio-IR correlation of \citet{bell03}. As in
  Fig.~\ref{fig:24um}, the blue and red regions denote the location of
  the models with maximum SFRs $\leq 30$ and $> 100$~\sfr,
  respectively. As these radio fluxes were derived from the starburst
  SEDs, they should also be treated as upper-limits. The dashed
  histogram is an estimate of the nuclear radio-emission from the
  central AGN assuming it is radio quiet, has a flat spectrum of
  $\alpha=0$ and $\nu L_{\nu}(5\ \mathrm{GHz})=10^{-5}L_{2-10\
  \mathrm{keV}}$ \citep{tw03}.}
\label{fig:radio}
\end{figure}

Another observational consequence of the nuclear starburst disks will
be radio emission. There have been numerous attempts to link the
properties of AGN found in deep radio and X-ray surveys
\citep{bauer02,georg04,simp06,bar07,richard07,rov07,park08}, but the
matched sources seems to only comprise a
small fraction of all AGN \citep[e.g.,][]{bar07}. Radio fluxes at 1.4~GHz can be
predicted for the starburst disk by making use of the IR SEDs and the
radio-IR correlation\footnote{The assumption that such intense
  starbursts will follow the observed radio-IR correlations may be
  problematic, as this might require very large magnetic energy
  densities \citep[e.g.,][]{thom06}. This problem will be investigated in
  future work.} \citep[e.g.,][]{yrc01}. We specifically use the form of the
radio-IR correlation derived by \citet{bell03} that employs the total
$8$--$1000$~\micron\ luminosity to make sure we account for the
emission from the pc-scale starburst. Observations by \citet{apple04}
and \citet{ibar08} have
shown that the local radio-IR correlation does seem to hold at high
$z$ (but see \citealt{bes08}). Therefore, upon assuming a typical starburst radio spectrum of
$\alpha=0.8$ (\citealt{yrc01}, where
$S_{\nu} \propto \nu^{-\alpha}$), we plot in Figure~\ref{fig:radio}
the distribution of radio fluxes from the `successful' starburst disks
at a representative redshift of $z=0.8$.
As in the previous figure showing the 24\micron\ predictions, the
contribution from models with a maximum SFR $\leq 30$ and $>
100$~\sfr\ are shown by the blue and red hatched regions,
respectively. The radio flux estimates are again upper-limits as they
originate from the starburst IR SEDs. The most common radio flux
predicted for these starburst disks is $\sim 10$--$30$~\microjy,
values that will again be dominated by disks with maximum SFRs $\leq
30$~\sfr. This is an interesting flux because it is just below the
sensitivity limit ($\sim 40$--$50$~\microjy) of the deepest radio
surveys \citep[e.g.,][]{bauer02,bar07}. Indeed, \citet{bauer02} found that the fraction of X-ray AGN
with radio matches increased as they approached their sensitivity
limit, and suggested that deeper observations will find a higher
fraction of matches. This conclusion agrees with the predictions of
Fig.~\ref{fig:radio}.

Interestingly, \citet{georg04} and \citet{richard07} find a slight tendency for the
fraction of radio/X-ray matches to increase with X-ray
obscuration. That is, the more obscured an AGN, the more likely that
it will have a radio counterpart in the deep radio surveys. This
result, although still tentative \citep{rov07}, is consistent with our proposed
scenario where obscured AGN should be associated with significant
star-formation. Future work is needed in the modeling to connect the
starburst disks to predictions of X-ray column densities. Further deep
radio observations are also required to confirm this trend.

As with the 24\micron\ flux, a major observational difficulty is
determining whether the origin of the radio flux is due to the AGN or to
star-formation. It has long been known that the radio source
population seems to change at a flux of $\sim 1$~mJy, with radio-loud
AGN dominating the counts above this flux, and, it was originally
thought, star-forming galaxies dominating below this flux
\citep{con84,wind85,shg04}. However, as suggested by \citet{jr04},
radio quiet AGN will start to contribute at faint radio fluxes, and,
indeed, recent observations find that AGN contribute $\sim 50$\% of
the radio population at the $\sim 50$~\microjy\ level \citep{simp06,sey08,smo08}. To estimate the radio-flux due to the nuclear emission
from the AGN, we use the result of \citet{tw03} who showed that for local
radio-quiet Seyfert galaxies and quasars, $\nu L_{\nu}(5\
\mathrm{GHz})=10^{-5}L_{2-10\ \mathrm{keV}}$. Combining this result
with the upper-limits for the X-ray luminosities plotted in
Fig.~\ref{fig:lx} results in a distribution of radio flux due to the
AGN itself (dashed histogram in Fig.~\ref{fig:radio}). To calculate the
1.4~GHz flux, a flat $\alpha=0$ spectrum was assumed, typical of the core
nuclear emission from an AGN \citep{ko88}. We find that the AGN radio flux at
$z=0.8$ are predicted to be only a factor of $\sim 3$ lower than the
starburst flux, indicating that confusion between AGN and
star-formation will be a factor in studying this population. However,
if the data are available, methods
such as spectral indices, radio morphology and radio-to-mid-IR flux
ratios have been shown to successfully discriminate between the two
possible origins \citep[e.g.,][]{don05,sey08}.

\begin{figure}
\epsscale{1.2}
\plotone{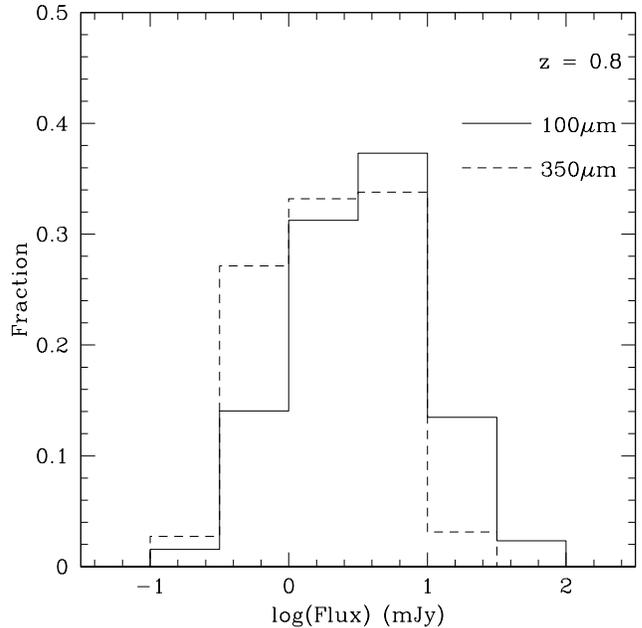}
\caption{The solid histogram plots the distribution of 100\micron\
  flux predicted from the starburst disk models at $z=0.8$. The dashed
  line plots the same results at 350~\micron.}
\label{fig:herschel}
\end{figure}

As noted above, both 24\micron\ and 1.4~GHz observations of distant
AGN will have difficulty distinguishing the presence of a nuclear
star-forming disk from the AGN emission. The starburst SEDs show a
wide range of emission because the star-formation occurs over a broad
range of temperatures. Therefore, it is useful to predict how these
starbursts disks may appear in the far-IR, especially with the
imminent launch of the \textit{Herschel Space
  Observatory}. Figure~\ref{fig:herschel} plots the 100\micron\ and
350\micron\ flux distributions from
our starburst models at $z=0.8$.
 Again, since these predictions are derived from the SED calculated
 using eq.~\ref{eq:sed}, they should be considered as
 upper-limits. Both of these distributions predict fluxes $\sim
 1$--$10$~mJy. The models of \citet{bal06b}
 were employed to estimate the contribution from an AGN heated torus at these
 wavelengths. At both these wavelengths, the torus emission was
 predicted to be over an order of magnitude smaller. Those
 calculations assumed a relatively compact geometry for the AGN
 absorber, it is possible a more extended Compton-thick absorber could
 produce greater far-IR emission, but we defer a calculation of such a
 scenario until later work.

Two \textit{Herschel} key projects are the PACS
Evolutionary Probe\footnote{http://www.mpe.mpg.de/ir/Research/PEP/} that has a planned sensitivity limit at
100\micron\ of 1--6~mJy. At 350\micron\ the Herschel Multi-tiered Extragalactic
Survey\footnote{http://astronomy.sussex.ac.uk/$\sim$sjo/Hermes/} (HerMES) predicts a sensitivity of 10--20 mJy. Thus, we predict
that the PACS Evolutionary Probe survey should detect these starburst
disks associated with hard X-ray sources. The HerMES survey will
likely only detect the very rare high star-forming or low-redshift
sources.

To summarize, the nuclear starburst disks proposed in this paper may
reveal themselves to observers in multiple ways. They should have
already been detected in the mid-IR by \spitzer\ surveys, but the
predicted fluxes are of the same order as those from AGN heating of a
non-star-forming torus, leading to the familiar IR spectral
decomposition problem. At 1.4~GHz, the starburst disks are predicted
to be slightly fainter than the deepest current radio surveys. Thus,
as already indicated from previous work \citep{bauer02}, the fraction of
radio/X-ray matches should increase as surveys probe to fainter flux
levels. Unfortunately, the predicted radio flux levels are comparable
to what is expected from the central radio-quiet AGN. Determining the
radio spectral index or radio-to-IR flux ratios are two promising
techniques that have been shown to distinguish between star-formation
and nuclear activity in radio sources \citep[e.g.,][]{sey08}. Finally, the cleanest
test may occur in the far-IR, where the contamination from AGN heating
is expected to be minimized. Figure~\ref{fig:herschel} shows that these disks
should be easily detectable by the planned \textit{Herschel} surveys
at 100 \micron. The question then arises: if star-formation is
detected in these AGN host galaxies, is there a way to determine if it
is originating in the nucleus? We address this question in the next
section.

\section{Discussion}
\label{sect:discuss}
\subsection{Distinguishing Between Nuclear and Galactic Starbursts}
\label{sub:galactic}
In the previous section, some of the observational properties of the
nuclear starburst disks were explored, and we showed that they should
be detectable at both radio and IR wavelengths. However, at the
redshifts of interest ($z \sim 1$), it is more than likely that
star-formation will be ongoing at galactic scales within the AGN host
galaxy. Thus, unless the galaxy is resolved on a kpc scale or less, any
observational detection of star-formation signatures from an AGN host
galaxy may be ambiguous with regard to its location.

Yet, there is a key difference between the nuclear starbursts
presented here and the more usual star-formation that is ongoing on
large scales. As described in Sections~\ref{sub:review}
and~\ref{sub:physical}, the crucial pc-scale burst of star-formation
only occurs when the central temperature of the disk exceeds the dust
sublimation temperature. This region of the disk does not reach these
temperatures by heating by an AGN, but rather
as result of the increasing density required by a star-forming $Q=1$ disk in a
black hole potential. Thus, in contrast to typical Galactic star-forming
regions, these models predict significant
star-formation from a region with very hot dust emission, but not AGN
heated dust. Therefore, if such nuclear starbursts are common, there
should be more near-IR or 24\micron\ flux than expected from just pure
AGN heating. As mentioned above, if this is true then it may
complicate AGN selection based on power-law like or hot mid-IR colors
\citep[e.g.,][]{don07}.

\begin{figure}
\epsscale{1.2}
\plotone{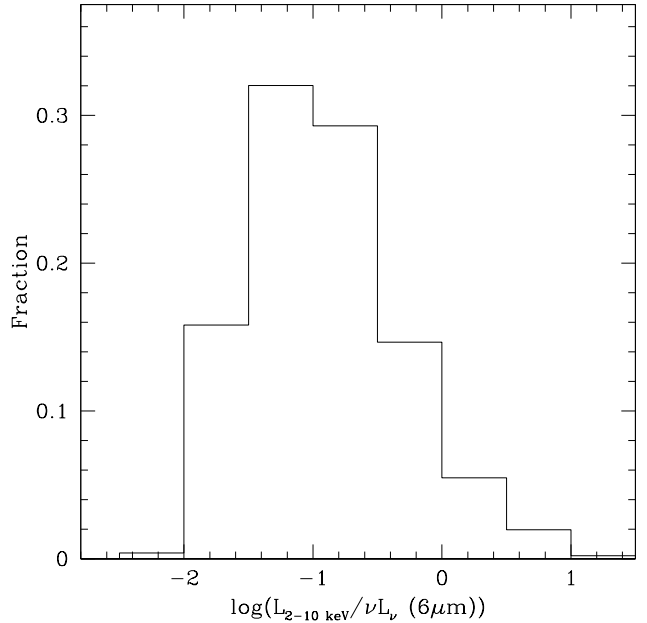}
\caption{Histogram of the estimated $L_{2-10\
  \mathrm{keV}}/\nu L_{\nu} (6\ \mu \mathrm{m})$ ratio predicted for
  the nuclear starburst models. They peak in the exact same range as
  the observations from both local and $z \sim 1$ AGN \citep{lutz04,rig06}.}
\label{fig:ratio}
\end{figure}

There may be evidence for this effect already in the
literature. Several observations of both nearby and high-redshift
AGN have shown that
the ratio of the absorption-corrected $L_{X}$ to a near- or mid-IR luminosity
is relatively constant with column density \nh\ \citep{lutz04,rig06,horst08}. This is not
expected from simple AGN unification and torus models where the
observed mid-IR luminosity is predicted to depend significantly on
\nh\ \citep{pk93}. Evidently, there is still hot dust emission observed from AGN at
all column densities. This fact has been used to argue that the AGN
torus should be clumpy \citep{horst08}, but it is also consistent with a nuclear
starburst disk around the AGN. To show this we plot in
Figure~\ref{fig:ratio} the distribution of the $L_{2-10\
  \mathrm{keV}}/\nu L_{\nu} (6\ \mu \mathrm{m})$ ratio for the
starburst models.
This histogram should only be taken as indicative, since both the
X-ray and IR luminosities are upper-limits. However, taking the
results at face-value, we find that the peak of the distribution is
very similar to the values observed by \citet{lutz04} and \citet{rig06}. Interestingly,
the AGN torus models of \citet{bal06b} were not able to reproduce
these ratios, as they were not able to produce a sufficient near-IR luminosity. While \citet{lutz04} argued that
they corrected their data for star-formation based on PAH emission,
if the starburst is produced in these hot pc-scale regions, the PAH
emission may be overwhelmed \citep{desai07} and the star-formation rate
underestimated. Thus, nuclear star-formation may be another mechanism
to explain the ubiquitous presence of hot dust emission even at large
column densities.

\subsection{Obscuration as a Function of AGN Luminosity}
\label{sub:obscuration}
Another interesting aspect of AGN phenomenology is the observed
decrease in the fraction of obscured AGN with increasing AGN
luminosity \citep{ueda03,laf05,ak06,tkd08}. This observation has been most commonly explained
by the receding torus model \citep{law91,simp05} which argues that the expansion of
the dust sublimation radius with luminosity will push back the torus
to larger radii and therefore decrease its covering factor. In this
paper, we propose that the AGN obscuration is dominated by a nuclear
starburst at a distance of $\sim 1$~pc. In this section, we investigate
if this model can also provide an explanation for the variation of the
Type 2/Type 1 ratio with luminosity.

\begin{figure}
\begin{center}
\includegraphics[angle=-90,width=0.50\textwidth]{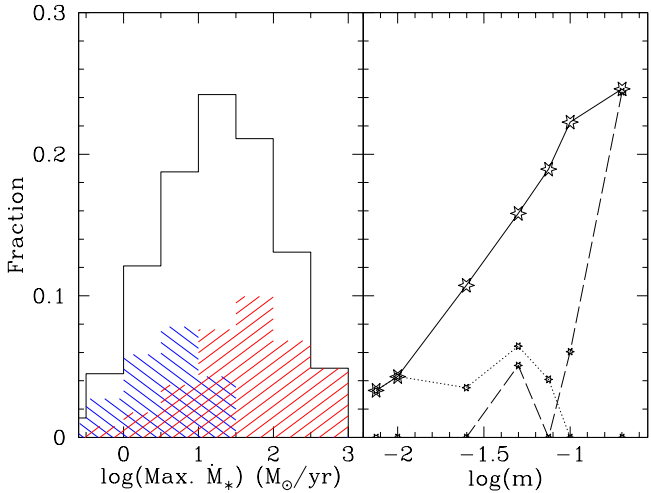}
\end{center}
\caption{(Left) The solid histogram plots the distribution of maximum
  SFRs seen in the models (as seen in Fig.~\ref{fig:sfr}). The blue
  and and red hatched regions denote the contributions from models
  with $42 < \log(L_X/\mathrm{erg\ s^{-1}}) < 43.5$ and $44 < \log(L_X/\mathrm{erg\ s^{-1}}) < 44.5$,
  respectively. (Right) The solid line plots the distribution of the
  angular momentum parameter $m$ found in the `successful' starburst
  disk models (as seen in Fig.~\ref{fig:props}(c)). The dotted and
  dashed lines denote the contributions from models
  with $42 < \log(L_X/\mathrm{erg\ s^{-1}}) < 43.5$ and $44 < \log(L_X/\mathrm{erg\ s^{-1}}) < 44.5$,
  respectively.}
\label{fig:vslx}
\end{figure}

Figure~\ref{fig:vslx} plots how two of the starburst parameters depend
on the estimated X-ray luminosity.
The left panel of the figure plots the distribution of the maximum
SFRs found in the model starburst disks. The blue and red hatched
regions denote the contributions from models with $42 < \log(L_X/\mathrm{erg\ s^{-1}}) <
43.5$ and $44 < \log(L_X/\mathrm{erg\ s^{-1}}) < 44.5$, respectively. This plot shows that
the most luminous AGN are associated with the most intense
star-formation, consistent with recent observational suggestions
\citep{shi07, wki08}. In contrast, the lowest luminosity AGN
are predicted with starburst disks with only modest star-formation
rates. The right panel of Figure~\ref{fig:vslx} plots the distribution
of the input angular momentum parameter $m$. In this case, the dotted
and dashed line denote the contributions from models
  with $42 < \log(L_X/\mathrm{erg\ s^{-1}}) < 43.5$ and $44 < \log(L_X/\mathrm{erg\ s^{-1}}) < 44.5$,
  respectively. As with the maximum SFRs, these two ranges of X-ray
  luminosities are separated in the diagram with large values of $m$
  only associated with luminous AGN. Similarly, starburst disks with
  weak angular momentum transport (i.e., low values of $m$) seem
  to only be able to feed a low-luminosity AGN.

These results can be used to posit an explanation for the observed
decrease in AGN obscuration fraction with luminosity. Recall that in
this starburst disk model there is a competition between
star-formation and gas accretion through the disk. If there is intense
star-formation within the disk then this reduces the gas supply
available to fuel a black hole. A circumnuclear starburst will
therefore more likely be associated with both a lower-luminosity and
obscured AGN. As the luminosity of the AGN increases, it is necessary
to also increase the gas supply onto the black hole. This could be
done, for example, by more efficient angular momentum transport in the
starburst disk (Fig.~\ref{fig:vslx}). While higher luminosity AGN can
be accompanied by very intense nuclear starbursts that may also
obscure the AGN, such bursts will likely be short-lived. More luminous
AGN, such as quasars, are unlikely to be fueled through these
starburst disks, because circumnuclear star-formation would consume
the necessary gas supply. The obscuration responsible for Type 2
quasars would then be generated at larger, galactic scales as a result
of the powerful events shaping the host galaxy \citep[e.g.,][]{hop05}.

This starburst scenario has an advantage over alternative mechanisms
that rely on AGN radiation to produce an obscuring medium
\citep[e.g.,][]{kro07,cqm07}. In these situations, more obscuration is
predicted with a larger AGN luminosity, as the vertical pressure
support of the obscuring medium is directly related to the intercepted
AGN luminosity. A significant outflow will need to develop in order to
limit the covering factor.  While it is clear that a more detailed
model of nuclear starburst disks is required to make quantitative
predictions, we find that, at least qualitatively, identifying a
pc-scale starburst as the origin of the obscuration for some AGN can provide a
natural explanation for the observed decrease in the fraction of
Type~2 AGN with luminosity.

\subsection{Torques and Angular Momentum Transport}
\label{sub:torques}
One of the most important parameters in determining whether a
circumnuclear starburst can both feed and obscure an active nucleus is
the efficiency of angular momentum transport through the disk. In this
paper we have assumed that a global torque, such as a bar or spiral
waves, operates on the starburst and removes angular momentum. We
further assume that the resulting accretion rate through the disk can
be sustained at a Mach number $m$. The results of
Sect.~\ref{sect:results} show that the likelihood for the formation of
a pc-scale starburst that could obscure an AGN is very sensitive to
$m$, with values of $0.1$--$0.2$ being most favorable, but lower
values are certainly possible, especially for lower-luminosity AGN. \citet{good03}
argues that spiral waves in an accretion disk could cause inflow with
$m \sim 0.1$, but it is not clear whether or not such global
coherent structures will be common for these disks. Some insight may
be gleaned from the study of gas flows in galactic nuclei (see \citealt{wada04} for
a recent review). High-resolution simulations of the gas dynamics in a
nuclear potential are now regularly finding coherent structures such as
bars or spirals in the gas flows \citep{es04,mac04}. These structures do remove
angular momentum and funnel material to the galactic center, although
the Mach number is not specified. These
gas flows may ultimately become turbulent which, as argued by \citet{wn02}, may
also cause further gas accretion.

Alternatively, a local viscosity mechanism, such as the
magneto-rotational instability \citep{bh98}, may operate in the starburst
disk. To determine how this different angular transport process might
affect the development of a pc-scale starburst, calculations of
starburst disks were run with a simple $\alpha$ viscosity prescription
(\citealt{ss73}; see also \citealt{tqm05}). For simplicity, a value of
$\alpha=0.3$ was considered with all the other parameters
($M_{\mathrm{BH}}$, $R_{\mathrm{out}}$, $f_{\mathrm{gas}}$, and the
dust-to-gas factor) varying as before. Fifty percent of the 180 models
generated a pc-scale starbursts with the parameters exhibiting very
similar trends to the ones shown in Fig.~\ref{fig:props}. However, as
found by \citet{tqm05}, this local viscosity does not move material
inward fast enough, and the resulting mass accretion rate remaining to
fuel the black hole is much smaller in this scenario than in the
global torque models described above. In fact, over half of the
`successful' $\alpha$-disk models resulted in AGN with
$L_{\mathrm{bol}}/L_{\mathrm{Edd}} < 0.002$. This accretion rate is
far too low to fuel the obscured AGN at $z \sim 1$ discovered by the
deep X-ray surveys.

We conclude that these relatively simple models suggest that in order
for starburst disks to be a common source of AGN obscuration and
fueling, an efficient angular momentum transport mechanism must
operate, and a global structure, such as a bar or spiral arms, is most
likely needed to provide such a mechanism.

\subsection{Starburst Disks and Unobscured AGN}
\label{sub:unobscured}
The focus of this paper has been on the properties of model starburst
disks that may potentially both obscure and feed an AGN. Only 41\% of
the model disks satisfied the selection criterion (see
Sect.~\ref{sub:seyferts}). Here, we discuss the properties of the
remaining 59\% of models that did not meet the imposed conditions. By
construction, these 748 starbursts disks do not have a region where
the central temperature surpasses the dust sublimation temperature and
produces an inner parsec-scale starburst. Therefore, the photosphere
of these disks is always thin and will not obscure any central AGN.
  
The vast majority of this population of starburst disks are unable to
fuel a central black hole at rates necessary to produce a luminous
AGN. In fact, 54\% of these disks would only produce an AGN with
$L_{\mathrm{bol}}/L_{\mathrm{Edd}} < 10^{-3}$ and 75\% with
$L_{\mathrm{bol}}/L_{\mathrm{Edd}} < 10^{-2}$. These models are ones
where $\tau_{\ast} < \tau_{\mathrm{adv}}$ and so nearly all the gas in
the disk is converted to stars at distances $\sim
R_{\mathrm{out}}$. The SFRs are within an order of magnitude to those
plotted in Fig.~\ref{fig:sfr}, but are in the cold, outer regions of
the disk. Observationally, such systems would appear as nuclear
starbursts within an inactive galaxy.

Interestingly, 25\% of these `unsuccessful' starburst disks, or 188
overall, can fuel an AGN sufficient to result in
$L_{\mathrm{bol}}/L_{\mathrm{Edd}} > 10^{-2}$. These are models where
the central temperature rises toward small radii, but just does not
obtain the sublimation temperature. There is no pc-scale burst of
star-formation, but star-formation does occur at modest rates over the entire
extent of the disk. Observationally, these models would manifest
themselves as unobscured, or Type 1, AGN. However, this class of
starburst disks exists in a very small region of the explored parameter
space, indicating that very precise conditions are required in order
to result in $\tau_{\ast} \la \tau_{\mathrm{adv}}$. This implies that
if a starburst disk is a viable mechanism to fuel the central black
hole, an obscured AGN in more likely to be produced than an unobscured
one. This conclusion agrees with the observed ratios of obscured to
unobscured Seyfert-like AGN at $z \sim 1$ \citep[e.g.,][]{bar05,gch07}.

Collecting together the results of all the starburst disk models reveals a
possible AGN evolutionary scenario. The results of
Sect.~\ref{sub:physical} show that obscured accretion will most likely
occur at large values of $f_{\mathrm{gas}}$. For small values of
$f_{\mathrm{gas}}$, the most likely outcome of a starburst disk will
be to turn the gas into stars at large radii and an inactive central
black hole. Thus, galactic nuclei may spend a large fraction of their
history with small amounts of ongoing nuclear star-formation and a
dormant black hole. However, processes at larger radii associated with
the assembly of the host galaxy will stochastic send gas-rich material
toward the nucleus. This material would have a large
$f_{\mathrm{gas}}$ and, upon passing through the starburst disk, would
result in an obscured, luminous AGN. The gas supply would slowly be
eroded, dropping $f_{\mathrm{gas}}$ and removing the inner
parsec-scale starburst, revealing an unobscured AGN. Finally, the gas
fraction would drop to low enough levels to cut off the fueling of the black
hole, and the system would return to producing stars at large
radii until the remaining gas is exhausted. While speculative, this
scenario may be a common occurrence during a stochastic phase of galaxy
assembly.

\section{Conclusions}
\label{sect:concl}
In this paper we proposed that nuclear starburst disks may be an
important contributor to AGN obscuration, especially at $z \sim 1$
where there is a large population of obscured AGN. The analytic model of
starburst disks developed by \citet{tqm05} was used to explore the
properties of such disks and determine if they were a viable and
general method to obscure accreting black holes. We found that a range
of conditions can produce potentially AGN obscuring bursts of
star-formation on pc scales. These include a high dust-to-gas ratio
and a relatively small size scale of the disk, both of which seem to
be consistent with current constraints. The pc-scale bursts of
star-formation are different from traditional star-forming
regions since they are produced in a region with a temperature $\sim
1000$~K. This region is not heated by an AGN, but is a
result of the assumption of a $Q=1$ stable star-forming disk in a
black hole potential. Thus, the observational signature of these
nuclear star-forming regions will be additional warm/hot dust emission
above what is required by AGN heating. This effect may be an
explanation for the observed $L_{2-10\ \mathrm{keV}}/\nu L_{\nu} (6\
\mu \mathrm{m})$ ratios seen from both local and high-redshift AGN. In
addition, the competition between star-formation and gas accretion
results in a natural explanation for the decrease in the fraction of
obscured AGN with luminosity. As an AGN-obscuring nuclear starburst
will consume gas that may have been destined for accretion onto a
black hole, this model predicts that nuclear starbursts should be
closely associated with lower luminosity AGN. In contrast, high
luminosity AGN require a significant gas supply, so a nuclear
starburst would be an hindrance to fueling the black hole.

The observational signatures of these disks may be most easily found
in future deep radio and far-IR surveys. Assuming the radio-IR
correlations hold for the circumnuclear starbursts, the expected 1.4~GHz fluxes
from these disks are $\sim 10$~$\mu$Jy at the redshifts where the
fraction of obscured AGN is near its peak. The 100\micron\ fluxes are
predicted to be $\sim 1$--$10$~mJy, which will be detectable in future
\textit{Herschel} surveys.

Nuclear starbursts may prove to be a compelling method of studying the
symbiotic relationship between black holes and their host galaxies in
the era of rapid evolution of both populations.  While qualitatively
interesting, this model needs to be explored further to make robust
quantitative predictions. A proper estimation of the vertical
structure of the disk is required to estimate covering
fractions. Future work should also include incorporating the feedback
from the central AGN. Another important future question is to determine
the nature and observable properties of the starburst remnants.

\acknowledgments

A careful reading of the paper by the anonymous referee is gratefully
acknowledged. The author is indebted to T.\ Thompson for very helpful
discussions and advice. DRB is supported by the University of Arizona
Theoretical Astrophysics Program Prize Postdoctoral Fellowship. 

{}

\end{document}